\documentstyle[12pt,epsf]{article}

\newcommand{\mathbold}[1]{\mbox{\rm\bf #1}}

\newcommand{\uone}{\mbox{\rm U(1)}}
\newcommand{\utwo}{\mbox{\rm U(2)}}
\newcommand{\uthree}{\mbox{\rm U(3)}}
\newcommand{\sutwo}{\mbox{\rm SU(2)}}

\newcommand{\be}{\begin{equation}}
\newcommand{\ee}{\end{equation}}
\newcommand{\bea}{\begin{eqnarray}}
\newcommand{\eea}{\end{eqnarray}}

\newcommand{\gtrsim}
{\ \rlap{\raise 2pt\hbox{$>$}}{\lower 2pt \hbox{$\sim$}}\ }
\newcommand{\lessim}
{\ \rlap{\raise 2pt\hbox{$<$}}{\lower 2pt \hbox{$\sim$}}\ }

\newcommand{\np}[1]{Nucl. Phys. {\bf #1}}
\newcommand{\pl}[1]{Phys. Lett. {\bf #1}}
\newcommand{\pr}[1]{Phys. Rev. {\bf #1}}
\newcommand{\prl}[1]{Phys. Rev. Lett. {\bf #1}}

\newcommand{\ptp}[1]{Prog. Theor. Phys. {\bf #1}}

\makeatletter
\if@twoside
\else \oddsidemargin 00pt \evensidemargin 00pt
\fi
\let\@eqnsel = \hfil

\expandafter\ifx\csname mathrm\endcsname\relax\def\mathrm#1{{\rm #1}}\fi
\@ifundefined{mathrm}{\def\mathrm#1{{\rm #1}}}{\relax}

\makeatother

\def\calo{{\cal O}}

\def\ie{{\it i.e.}}

\def\msusy{m_{\rm S}}
\def\be{\begin{equation}}
\def\ee{\end{equation}}
\def\bea{\begin{eqnarray}}
\def\eea{\end{eqnarray}}
\def\identity{1 \hspace{-.085cm}{\rm l}}

\begin{document}
\begin{titlepage}
\begin{center}
\hfill CERN--TH/95--207\\
\hfill hep-ph/9507462
\end{center}
\vskip .03in
\center{\Large\bf Horizontal Symmetries for the}
\center{\Large\bf    Supersymmetric Flavor Problem}
 \vskip .6in
\center{{\sc Alex
Pomarol and Daniele Tommasini}}
\center{{\it Theory Division, CERN}\\{\it CH-1211 Geneva 23,
Switzerland}}
\vskip .6in
\begin{abstract}

The heaviness of the third family fermions and the experimental absence
of large flavor-violating processes suggest, in  supersymmetric theories,
that the three families belong to a ${\bf 2+1}$
representation of a horizontal symmetry $G_H$.
In this framework, we discuss a class of models, 
based on the group U(2), that
describe the flavor structure of the fermion masses and are compatible
with an underlying GUT.
We study the phenomenology of these models
and focus on two interesting scenarios:
In the first one, the  first and second family scalars are
assumed to be heavier than the weak scale,
allowing for complex soft supersymmetry-breaking terms.
In the second one, all the  CP-violating phases
are assumed to be small. Both scenarios present a rich phenomenology
 in agreement with  constraints from flavor-violating processes
 and electric dipole moments.

\vfill

\noindent CERN--TH/95--207\hfill\\
\noindent July 1995\hfill
\end{abstract}
\end{titlepage}

\section{Introduction}

The flavor structure of the Standard Model (SM) is an
open problem, but also a hint in the search for a more fundamental
theory. From the experimental data we have learned that
the third-family fermions are very special.
They are very heavy
and   almost decoupled
from the  fermions of the other two  families.
In the supersymmetric version of the SM (MSSM), data from
flavor-violating processes provide us with more hints
about the flavor structure.
The experimental indication of small
flavor changing neutral currents (FCNC)
 implies
the need of a super-GIM mechanism in the scalar-quark sector, \ie\
 the first- and second-family squarks have to be   highly
degenerate \cite{dg}.
To satisfy such a requirement,  the scalar masses 
were assumed to be universal
(universality condition) \cite{dg}.
Nevertheless,  if no symmetry protects such universality,
it will be spoiled by higher-order effects \cite{hkr}. Then, even
if universality holds at the scale  where supersymmetry is broken,
it will not hold at a lower scale.
This departure from universality induced by 
radiative corrections can be particularly significant in 
grand unified theories (GUTs) where large representations and large Yukawa
couplings are present \cite{hkr,pp}.

The above considerations suggest that any low-energy effective
supersymmetric theory
must have an approximate U(2)$_L\times $U(2)$^d_R\times $U(2)$^u_R$
symmetry of flavor under which the $1^{st}$ and $2^{nd}$ families
transform as a doublet and the $3^{rd}$ as a singlet.
Such flavor symmetry  could just be accidental.
In this paper, however,
we will consider   that
a horizontal symmetry $G_H$
with the three families belonging to the ${\bf 2+1}$
representation is actually realized
 at some high-energy scale.
Furthermore,
inspired by GUTs,
 we will consider that $G_H$
is a subgroup of only one U(2) factor instead of allowing for independent
left and right rotations\footnote{
Our results can be easily generalized to the case
 U(2)$_L\times $U(2)$^d_R\times $U(2)$^u_R$.}.
The $G_H$ symmetry  could be
gauged, global or discrete. If the horizontal group is gauged, however,
 gauge-breaking effects can arise in the squark sector
 (D-term contributions), which, as we will see, are too large.
This suggests that $G_H$ could be a
 discrete subgroup of U(2). 
(Moreover, in the latter case one does not need to worry about the 
cancellation of the U(1) anomaly). 

Of course, $G_H$ has to be broken to give mass to the
lightest families.
Motivated by the pattern of fermion masses and mixing angles,
we will consider that the horizontal symmetry breaking is realized
in two steps. In the first step,
  the second-family fermions get masses, while it is only after
 the second step
that the first family becomes massive.

We will not make any specific assumption about  the soft supersymmetry
breaking (SSB) terms.
Based on the above symmetry-breaking pattern,
we will be able to infer the flavor structure
of the  scalars mass matrices and analyse
the phenomenological implications of the model.
The horizontal symmetry will force an approximate
$1^{st}-2^{nd}$ family squark degeneracy,
and FCNC processes will be suppressed (by the super-GIM mechanism).
We will also study the possibility that the first two families of
scalars are heavier than the other particles
and allow for  complex SSB terms.
These two ingredients will lead to a phenomenology very
different from that in the MSSM with universal soft masses.

Horizontal symmetries under which the quarks transform as {\bf 2+1}
have been proposed previously in the literature to solve the
 supersymmetric flavor problem \cite{seiberg,dine} (see also
ref.~\cite{kaplan}).
Our approach, however, will be more general and address
new aspects, such as
the $G_H$-breaking effects in the SSB terms, 
the problems with gauging $G_H$,
CP violation, third-family contribution to FCNC, and
electric dipole moment (EDM) contributions.

In section 2, we discuss the problem of the quark masses and mixing angles
from a phenomenological point of view, and look for the possible
approximate horizontal symmetries $G_H$ of the SM that allow us to reproduce
the observations.
Considering in particular the non-Abelian case with the fermions in the
$\mathbold{2}+\mathbold{1}$ representation, we present
two toy models, one leading to a texture form for the quark mass matrices,
the other allowing for more general structures compatible with the 
symmetry-breaking pattern.
In section 3, we present the flavor structure of the scalar mass matrices
imposed by the horizontal symmetry breaking. We also discuss the
possibility that
 the first- and second-family squarks are heavier than the other scalars.
In section 4, we  analyse the phenomenology of the model. We study the
contribution to FCNC, EDMs, lepton-number-violating processes, proton
decay and the effects of the CP-violating phases in the
particle mass spectrum.
Section 5 is devoted to conclusions.
In the appendix, we show how the SSB terms can be  modified when the
horizontal symmetry is broken.

\section{Fermion masses and horizontal symmetries}

\subsection{General considerations}

In a purely phenomenological approach,
the observed quark masses and the Cabibbo-Kobayashi-Maskawa (CKM)
\cite{ckm}
mixing matrix $V_{\rm CKM}$ can be written in the Wolfenstein
parametrization \cite{wolfenstein} as
\be
M_d^{\rm diag}=m_b\pmatrix{d\lambda^4&0&0\cr
                         0&s\lambda^2&0\cr
                         0&0&1\cr},\qquad
M_u^{\rm diag}=m_t\pmatrix{u\lambda_u^4&0&0\cr
                         0&c\lambda_u^2&0\cr
                         0&0&1\cr},
\label{1.1}
\ee
\be
V_{\rm CKM}=
\pmatrix{1-\lambda^2/2            & \lambda    & A\lambda^3(\rho+i\eta)\cr
-\lambda                 & 1-\lambda^2/2 & A\lambda^2            \cr
A\lambda^3(1-\rho+i\eta) & -A\lambda^2   & 1                   \cr},
\label{1.2}
\ee
where $d,s,c\simeq1$, $u\lessim1$, $\lambda\simeq0.2$ is essentially the
Cabibbo angle, $\lambda_u\simeq0.06$, and all the coefficients
 in eq.~(\ref{1.2}) are of order 1.

In general, the above structure is not sufficient to determine the form of
the down- and up-quark mass matrices $M_d$ and $M_u$
in the basis of the gauge eigenstates.
However, the Cabibbo angle turns out to be given with a good
approximation by the same parameter $\lambda\simeq\sqrt{m_d/m_s}$ that
determines the matrix $M_d^{\rm diag}$, and
from $\lambda>\lambda_u$
we can expect that the full CKM matrix is given in a first
approximation by the (left-handed) down-quark rotation $U_L^d$,
which relates the gauge to the mass eigenstates.
This consideration is reflected in eqs.~(\ref{1.1}) and (\ref{1.2}), where
the same parameter $\lambda$ has been used to describe the down-quark
masses and the CKM mixing matrix.

Neglecting the up-quark rotation, and assuming for the moment
that $M_d$ is a symmetric matrix (so that $U_R^d=U_L^{d*}$),
we can guess the form of the down-quark mass matrix in the gauge
eigenstates basis
\be
M_d=U_R^d M_d^{\rm diag}U_L^{d\dagger}
\simeq V_{\rm CKM}^* M_d^{\rm diag} V_{\rm CKM}^\dagger\simeq\pmatrix{
(d+s)\lambda^4          & s\lambda^3        & A (\rho-i\eta)\lambda^3 \cr
s\lambda^3              & s\lambda^2        & A\lambda^2              \cr
A (\rho-i\eta)\lambda^3 & A\lambda^2        & 1\cr}
m_b,       \label{1.3}
\ee
at the lowest order in the expansion parameter $\lambda$.
Then, from these semi-phenomenological considerations,
we find the following `onion' structure for the mass matrix $M_d$:
$i$) the entry $33$ is of order $\lambda^0=1$;
$ii$) the  $22$, $23$ and $32$  entries appear at order
$\lambda^2$; $iii$) the off-diagonal elements of the first row (and column)
are of order $\lambda^3$; $iv$) the $11$ entry arises only at ${\cal
O}(\lambda^4)$. This structure is expected even if $M_d$ is not
symmetric, provided that $(U_R^d)_{ij}\sim (U_L^d)_{ij}^*$.
In this case, eq.~(\ref{1.3}) should be understood
as an order-of-magnitude estimate for the entries of $M_d$, with
undetermined coefficients of order 1 for each $\lambda^n$ term.

Equation (\ref{1.3}) suggests that a spontaneously broken
horizontal symmetry, acting on the flavor indices, could
be responsible for the generation of the quark masses. Assuming that the
parameter $\lambda$ measures the amount of the breaking of
the symmetry, we can guess that the subsequent `onion' layers
correspond to different breaking steps.
We anticipate that symmetry arguments alone cannot explain
the orders of magnitude of the ratios of the different
layers of eq.~(\ref{1.3}), and in particular they do not
predict the suppression in the $11$ entry of the
mass matrix $M_d$ since, 
 according to eq.~(\ref{1.3}), it should be a factor $\lambda$
smaller than the other elements in the first row (and column).
These problems can be solved in the class of models that we will discuss,
where the different powers in $\lambda$ appearing in eq.~(\ref{1.3}) will
be related to the order of the non-renormalizable operator contributing
to the given mass entry.
Since we do not necessarily require $M_u$ and $M_d$ to have texture zeros
\cite{texture}, we will
generally content ourselves to predict just the dependence of
the entries in eq.~(\ref{1.3}) on $\lambda$. In other words, we will
just try  to explain the orders of magnitude for the mixing angles, not the
precise values.
Nevertheless, we will also provide an example where a texture structure
is obtained, so that accurate predictions for the CKM mixing matrix
can be made, following e.g. ref.~\cite{texture}.

Once the CKM matrix is assumed to be essentially related to the down-quark
rotation, the flavor structure of the matrix $M_u$ remains undetermined.
However, if the horizontal symmetry acts in the same way on the up- and
down-quark sectors, then $M_u$ and $M_d$ can be expected to have similar
structures. In GUTs  such as SO(10),
the matrices $M_u$ and $M_d$ (more precisely, the Yukawa couplings at the
GUT scale) are related by Clebsch-Gordan coefficients \cite{adhrs},
which should take into account the fact that the scaling factor for the
up-quarks $\lambda_u$ is smaller than the factor $\lambda$ determining the
down-quark mass ratios.  We will not try to explain the
 large ratio $m_t/m_b$; it could arise from
a large $\tan\beta\equiv\langle H_u\rangle/\langle H_d\rangle$ or from
a small  coupling of the down sector with the light Higgs.

Motivated by the previous phenomenological considerations,
our aim here is to classify the approximate horizontal groups $G_H$ leading
to eq.~(\ref{1.3}).
In the limit of vanishing Yukawa couplings, the SM has a global flavor
symmetry $\uthree$ for each field in the basis of the left-handed fields
$Q=\pmatrix{u & d\cr}^T$, $u^c$, $d^c$, $L=\pmatrix{\nu & e\cr}^T$ and $e^c$
(with a flavor index understood, and
$\psi^c\equiv \psi^c_L\equiv C(\bar \psi_R)^T$).
This would be the maximal horizontal symmetry allowed
 for any effective theory,
containing no new fields in addition to the known SM (or MSSM) particles.
In the following we will assume that all the SM fermions in the basis given
above transform in the same way under $G_H$.
This is a necessary condition if $G_H$ is assumed to commute with some
GUT group such as SO(10).
In other words, we will not consider here the alternative 
that the horizontal symmetry is unified in a non-trivial way within
a larger GUT.

We assume that $G_H$ allows (just) the third family masses,
\ie\ mass matrices of the form
\be
M_d^0=M_d(\lambda=0)={\rm diag}(0,0,1)m_b
.\label{1.4}
\ee
Since we have assumed that the quark fields $Q$ and $d^c$
transform in the same way, then
$G_H $ is the group of (unitary) matrices $U$ satisfying the condition
\be
U^T M_d^0 U=M_d^0
\label{1.5}
,\ee
where we have also assumed that the Higgs field responsible for
the breaking of the SM gauge symmetry is invariant under $G_H$. This
latter simplifying assumption characterizes our approach,
which can be easily generalized to the case that the light 
Higgs field transforms as a singlet with a non-trivial phase under $G_H$. 
The most general solution $U\in\uthree$ of eq.~(\ref{1.5}) is a block matrix
$U=\pmatrix{U_{2\times2}&0_{2\times1}\cr
            0_{1\times2}&\pm1\cr} $,
that is
\be
G_H\subseteq \utwo\times Z_2=\uone\times \sutwo\times Z_2
,\label{1.6}
\ee
where the U(1) and SU(2) factors correspond
to global phase and special unitary
transformations for the first two families, and the $Z_2$ describes a sign
ambiguity in the definition of the third family, which
is singlet under $G_H$ (the factor $Z_2$ in eq.~(\ref{1.6}) 
would be  a U(1) if we 
 allowed the light Higgs field to transform with a 
 non-trivial phase under $G_H$). 
We are then left with two possibilities for the first two
families: they can belong either to A) two singlets under $G_H$,
so that the three-flavor components quark field belongs to the
representation ${\bf 1+1+1}$, or to B) a doublet, so that the quarks
belong to ${\bf 2+1}$.

The first case (A) results, for instance, if $G_H=\uone$ in eq.~(\ref{1.6}).
It is interesting to notice that this minimal continuous group, or even its
$Z_3$ subgroup (consisting of discrete phase transformations
 given by the third
roots of the unity), is 
sufficient to force the form eq.~(\ref{1.4}) of the mass matrix.
In fact, it can be shown that a theory based on $G_H=\uone'\times Z_2$,
where the $\uone'$ is generated by
a suitable combination of the U(1) and SU(2)
generators in
eq.~(\ref{1.6}), can provide mass matrices of the form (\ref{1.3}).
A similar result can be obtained using the $Z_8$ discrete subgroup of
$\uone'$ \footnote{Models based on a U(1) horizontal  symmetry
can also reproduce texture structures for the mass matrices \cite{u1}.}.
However, this case (A) is disfavored
in the supersymmetric version of the
theory, since it cannot provide a super-GIM mechanism
for the scalar sector.

\subsection{$G_H$ with the quarks in the ${\bf 2+1}$ representation}

In the case (B) mentioned in the previous paragraph, 
$G_H$ should be non-Abelian
in order to have a two-dimensional irreducible representation. From
eq.~(\ref{1.6}), it should then contain
either the SU(2) factor, or one of its (discrete) non-Abelian subgroups
having the same doublet representation, such as the
double dihedral groups $Q_{2N}\equiv D_N^D$ \cite{cornwell}, that have
$4N$ elements.
For definiteness, we will consider the cases $Q_{2N}$ with either $N=3$ or
$N=4$, since they reproduce all the properties of the continuous case for
what concerns the fermion masses in the cases we will
consider\footnote{These discrete groups were also used in ref.~\cite{frampton}
to predict texture forms for the quark masses in the framework of
non-unified theories.}.
We remark again that in the supersymmetric version of the theory the
case of the discrete symmetry will be selected as phenomenologically
preferred, since a gauge horizontal symmetry leads, in general, to
unacceptably large FCNC.

It is interesting to notice that the choice $G_H=\sutwo$ (or $G_H=Q_{2N}$)
is also sufficient by itself to ensure the form eq.~(\ref{1.4}) of $M_d^0$,
if the quark Yukawa couplings to
the SM Higgs field are symmetric in the flavor
indices. This is the case in SO(10) models where the light Higgs field
belongs to the (symmetric) {\bf 10} representation.
However, an additional $\uone$
factor is helpful to generate
acceptable mass matrices after the spontaneous horizontal symmetry breaking.
The corresponding factor  in the discrete group case is $Z_{2N}$,
which has $2N$ elements, as the $\simeq Z_{2N}$ Abelian subgroup of $Q_{2N}$,
which consists of the {\it diagonal} matrices \cite{cornwell}
\be
\left\{\pmatrix{e^{i\pi k/N} & 0 \cr
            0 & e^{-i\pi k/N} \cr}, \quad (k=0,...,2N-1)\right\}
.\label{1.7}
\ee

Let us study first what can be deduced on the quark masses from
symmetry arguments alone in models based on the full
group of eq.~(\ref{1.6}), or on its discrete subgroup $Z_{2N}\times Q_{2N}
\,(\times Z_2)$.
In the continuous case, we will consider the symmetry-breaking chain
\be
\uone\times\sutwo \to\uone_1 \to 1
,\label{1.71}
\ee
where the $\uone_1$ acts only on the first flavor\footnote{Taking into
account the possible additional $Z_2$ factor, the symmetry breaking chain is
$\uone\times\sutwo\times Z_2 \to\uone_1\times Z_2'\to Z_2^{\prime\prime}$,
where
$Z_2'$ describes a simultaneous change of the second and third families
indices, and $Z_2^{\prime\prime}$ is the global sign ambiguity.}.
The corresponding symmetry-breaking pattern in the discrete case is
\be
Z_{2N}\times Q_{2N}\to Z'_{2N}\to 1
,\label{1.72}
\ee
where $Z'_{2N}$ corresponds to the discrete
phases for the first family (\ie\ it is the discrete $Z_{2N}$
subgroup of $\uone_1$),
and can be thought to result from the combinations of the $2N$ elements
of $Z_{2N}$ with the $2N$ diagonal matrices of $Q_{2N}$,
 eq.~(\ref{1.7}). [Actually the $Z_{2N}'$ acts as a $Z_N'$ on the doublet 
quark field, since it is represented by the matrices 
$\pmatrix{e^{2i\pi k/N} & 0 \cr
            0 & 1 \cr}$.] 
The first symmetry-breaking step in eq.~(\ref{1.71})
 [or in eq.~(\ref{1.72})]
 generates non-zero 22, 23 and 32 entries in the mass matrix $M_d$,
giving rise to the second family masses and the CKM mixing angle $V_{ts}$
(and $V_{cb}$). If the 22, 23 and 32 entries are of the same order of
magnitude, then
$V_{ts}\sim m_s/m_b$, which is in the experimentally correct range.
Since the $\uone_1$ (or $Z'_{2N}$) symmetry protects the
first row and column of the mass matrix, the first family mass and its
mixing angles can only be generated after it is
broken. In general, the
11 entry is expected to be of the same order as the 12, 13, 21 and 31
ones, in contradiction with the `onion' pattern discussed in the previous
section.
Nevertheless, if the SU(2) (or $Q_{2N}$) symmetry is broken
only by the VEVs of $G_H$-doublet scalar fields, the 11 entry turns out to be
proportional to the square of the $\uone_1$ (or $Z'_{2N}$)
breaking parameter and
can be conveniently suppressed, as we will see explicitly below.

Let $G_H=\uone\times\sutwo\times Z_2$
(or $G_H=Z_{2N}\times Q_{2N}\times Z_2$),
allowing the Yukawa Lagrangian
\be
{\cal L}^0_{\rm Y}=h_d d^c_3 H_d Q_3 + h_u u^c_3 H_u Q_3+h.c.,
\label{1.73}
\ee
where for each term we have understood a Charge Conjugation
matrix, and $H_d$ ($H_u$) is the MSSM Higgs field giving mass to the down-
(up-) quarks.
The quark-mass terms involving the first two families
will arise from non-renormalizable operators involving the scalar fields
whose VEVs break $G_H$. These contributions are suppressed by inverse powers
of the energy scale $\Lambda$, the cutoff of our effective theory,
which can be close to the GUT scale $M_G$.
We then assume that the first breaking step in eq.~(\ref{1.71}) 
[eq.~(\ref{1.72})] occurs at some scale below $\Lambda$,
due to an $\sutwo$-doublet scalar field $\Phi$, which is
singlet under the SM group (see table 1)\footnote{If the underlying 
GUT  is SO(10),
these fields should actually belong to non-trivial representations
such as the $\mathbold{45}$, in order to
 generate different $M_u$ and $M_d$ matrices \cite{adhrs}.}.
Without losing generality, it is always possible to rotate to a
basis in which the first component of $
\langle\Phi\rangle$ vanishes. Then we can write
\be
\langle\Phi\rangle=
\Lambda\pmatrix{0  \cr
                              \epsilon \cr},
\label{1.8}
\ee
where $\epsilon$ is a small adimensional parameter 
(we are assuming that $i\sigma_2\Phi$ transforms as the quark 
$G_H$-doublet); $\langle\Phi\rangle$ preserves the residual
$\uone_1$ (or $Z'_{2N}$) symmetry of eq.~(\ref{1.71}) [or (\ref{1.72})],
acting on the first family\footnote{The rotation leading to eq.~(\ref{1.8})
does not belong in general to the discrete group $Q_{2N}$, and we should
assume that the $Z'_{2N}$ symmetry is left invariant by $
\langle\Phi\rangle$. Actually, in the models discussed below
the hierarchy amongst the entries in $M_d$ can be due to an increasing
negative power in $\Lambda$ rather than to
a hierarchy between the $Q_{2N}$ and the $Z'_{2N}$ symmetry-breaking
scales.},
which is generated in this basis by the charge ${\cal Q}_1=T_3+{\cal Q}/2$
(where $T_3$ and ${\cal Q}$ are the SU(2) and U(1) generators).
Then non-renormalizable terms involving $(\langle\Phi\rangle/\Lambda)^2$
will be responsible for the generation of the 22 entry of
the mass matrix at the order $\epsilon^2$, while
linear terms in $\langle\Phi\rangle/\Lambda$ will contribute to the 23 and 32
entries. According to the `onion' structure of eq.~(\ref{1.3}),
all these entries should be of
the same order of magnitude, so that the 23 and 32 elements should
also be suppressed by an extra power in $\epsilon$.
This can be guaranteed by the symmetry $Z_2$ if
a $\uone\times\sutwo$ (or $Z_{2N}\times Q_{2N}$)
invariant field $\chi$ is introduced, and
the transformation rules of table 1 are assumed.
The field $\chi$ is supposed to get a $Z_2$-breaking VEV, which can be
written in terms of an adimensional parameter $\epsilon_\chi$ as
\be
\langle\chi\rangle=\Lambda \epsilon_\chi
\label{1.8c}
.\ee
\begin{table}[ht]
\begin{center}
\begin{tabular}{|c||c|c|c|}
\hline
         & $\uone$ (or $Z_{2N}$) & $\sutwo$ (or $Q_{2N}$) & $Z_2$ \\
\hline
\hline
$q\equiv\pmatrix{q_1\cr q_2\cr}$ & 1 & $\mathbold{2}$ & +1 \\
$q_3$                            & 0 & $\mathbold{1}$ & $z(=\pm1)$  \\
\hline
$\chi$                           & 0 & $\mathbold{1}$ & $-1$ \\
$\Phi\equiv\pmatrix{\Phi^1\cr \Phi^2\cr}$ &
                                $-1$ & $\bar{ \mathbold{2}}$ & $-z$ \\
\hline
I) $\Phi'\equiv\pmatrix{\Phi^{\prime1}\cr \Phi^{\prime2}\cr}$ &
                                $-1$ & $\bar{ \mathbold{2}}$ & $-z$ \\
\phantom{$\chi_1$ }
II)
$\chi_1$ \phantom{$\Phi'\equiv\pmatrix{\Phi^{\prime1}\cr \Phi^{\prime2}\cr}$ }
                                  & $-{2/3}$ & $\mathbold{1}$ & +1\\
\hline
\end{tabular}
\vskip 2cm
\caption{Horizontal symmetry assignments for the fields in a
model based either on the group $\uone\times\sutwo\times Z_2$,
or on the discrete symmetry $Z_{2N}\times Q_{2N}\times Z_2$.
The $\uone$-generator ${\cal Q}$ can  also be used to define the
representation of $Z_{2N}$, except in the case of the
field $\chi_1$, which should transform under $Z_8$ as the quark field
squared.
The label $q_i$ refers to any of the quark fields, $q_i=Q_i,d^c_i,u^c_i$.}
\end{center}
\end{table}

The $G_H$-symmetric Yukawa
Lagrangian, including the lowest-dimension non-renormalizable interactions,
can be written as\footnote{The discrete symmetry $Z_{2N}\times Q_{2N}$ allows
for few more invariant terms, such as
$d^{cT}\sigma_1\sigma_2 \Phi\Phi^T\sigma_2\sigma_1 Q$ and
$d^{cT}\sigma_2 \Phi\Phi^T\sigma_2 Q-
d^{cT}\sigma_3\sigma_2 \Phi\Phi^T\sigma_2\sigma_3 Q$. It can be seen
that the form of the mass matrices below is not changed by such additional
invariants, which affect only the actual coefficients of the different
entries.}
\be
{\cal L}_{\rm Y}= h_d
\pmatrix{d^{cT} & d^c_3\cr}
\pmatrix{
{h_{\Phi\Phi}^d\over \Lambda^2} \Phi\Phi^T &
{h_{\Phi\chi}^d\over \Lambda^2} \Phi\chi \cr
{h_{\chi\Phi}^d\over \Lambda^2} \Phi^T\chi &
1 \cr}
\pmatrix{Q\cr Q_3\cr} H_d +\{d\to u\} + h.c.
,\label{1.9}
\ee
where $h_{\Phi\Phi}^d$, $h_{\Phi\chi}^d$ and $h_{\chi\Phi}^d$ are Yukawa
couplings normalized to $h_d$, which can be expected to be of
${\cal O}(1)$. As mentioned above, in GUTs such as SO(10)
it has to be assumed that the difference between
the up-quark and down-quark Yukawa couplings is due to different
Clebsch-Gordan factors that depend on the  GUT.
When the electroweak symmetry is broken by $\langle H_d\rangle$,
the mass matrix
\be
M_d=\pmatrix{
0 & 0 & 0 \cr
0 & h_{\Phi\Phi}^d\epsilon^2 & h_{\Phi\chi}^d\epsilon\epsilon_\chi \cr
0 & h_{\chi\Phi}^d\epsilon\epsilon_\chi & 1         \cr}
h_d\langle H_d\rangle
\label{1.10a}
\ee
is generated,
and we will assume that $\epsilon_\chi\sim\epsilon$, which is natural if
the horizontal $\sutwo$ and $Z_2$ symmetries are broken at a common scale.
Now the identification
$\epsilon\sim\lambda$ will provide the correct orders of magnitude for the
 22, 23 and 32 entries of $M_d$, as in eq.~(\ref{1.3}).

For the first row and column of the mass matrix to be generated,
the $\uone_1$ (or $Z_{2N}'$) symmetry has to be broken. As shown in table 1,
this can be due to either a second $\sutwo$-doublet $\Phi'$ (case I),
or  II) an $\sutwo$-singlet $\chi_1$ (case II).

{\bf Model I}:
The addition of the second $\sutwo$-doublet $\Phi'$, getting the general VEV
\be
\langle\Phi'\rangle= \Lambda\pmatrix{\delta'\cr
                                     \epsilon'\cr},
\label{1.8b}
\ee
allows us to generate  the first family mass and its mixing angles.
Omitting coefficients of order 1, the full mass matrix now reads 
\be
M_d\simeq\pmatrix{
\delta^{\prime 2} & \epsilon'\delta' + \epsilon\delta'
& \delta'\epsilon_\chi\cr
\epsilon'\delta' + \epsilon\delta' & \epsilon^2+\epsilon^{\prime2}+
                            \epsilon\epsilon'&
          \epsilon\epsilon_\chi + \epsilon'\epsilon_\chi\cr
\delta'\epsilon_\chi & \epsilon\epsilon_\chi +\epsilon'\epsilon_\chi & 1
\cr}
h_d\langle H_d\rangle
.\label{1.10}
\ee
If the $\uone_1$-breaking parameter $\delta'$ is of the order
of $\epsilon^2$,
eq.~(\ref{1.10}) reproduces the `onion' pattern of eq.~(\ref{1.3}).
This condition is satisfied naturally if $\Phi'$ is an effective
field given by the product of two fields
(in this case we also expect $\epsilon'\sim\epsilon^2$).
For instance, we can consider a model containing
a doublet $\Phi^{\prime\prime}$
having ${\cal Q}=-2$ (and $Z_2$-parity $=-z$),
and an SU(2)-singlet $\chi^{\prime\prime}$  ($Z_2$-even)
carrying ${\cal Q}=1$. Then we can define 
$\Phi'\equiv\Phi^{\prime\prime}\chi^{\prime\prime}$, and we have 
\be
\delta',\epsilon'\sim
{\langle\Phi^{\prime\prime}\rangle\langle\chi^{\prime\prime}\rangle
\over\Lambda^2}
\sim\epsilon^2
\label{1.12b}
,\ee
assuming that all the $G_H$-breaking VEVs are generated at the same
energy scale.
Then eq.~(\ref{1.10}) reproduces the structure of eq.~(\ref{1.3})
if the common order of magnitude for the adimensional parameters
$\epsilon$ and $\epsilon_\chi$ is
$\lambda$.

We remark that in this example the mass matrix eq.~(\ref{1.10})
is not necessarily symmetric in the flavor indices, 
so that the comparison with eq.~(\ref{1.3})
cannot be strict. In fact, in SO(10)$\times G_H$ models
the fields $\Phi$ and $\chi$ can belong to the
$\mathbold{45}$ representation of SO(10).
As a consequence, the mass matrix is in general not symmetric,
since the product $\mathbold{45}\times\mathbold{10}$ contains
the $\mathbold{120}$ {\it antisymmetric} representation. However,
we expect from eq.~(\ref{1.10}) that in orders of magnitude
$(M_d)_{ij}\sim (M_d)_{ji}$, so that
$U^d_L\sim U^{d*}_R$.
The precise coefficients in eq.~(\ref{1.10})
are related to the ratios of different Yukawa couplings, which are
arbitrary numbers expected to be of order 1. For this reason, in model I
there is no hope to predict more than the orders of magnitude for the CKM
mixing angles as functions of the quark masses.

{\bf Model II}: In this case, the $\uone_1$ (or $Z'_{2N}$)
symmetry is broken by the VEV
of an SU(2)-singlet field $\chi_1$, as given in table 1, which can be written
in terms of an adimensional parameter $\epsilon_1$ as
\be
\langle\chi_1\rangle= \Lambda \epsilon_1
.\label{1.13}
\ee
Then the non-renormalizable Yukawa Lagrangian eq.~(\ref{1.9}) gets
an additional operator involving the antisymmetric SU(2) invariant
$d^{cT}i\sigma_2 Q=d^c_1 Q_2-d^c_2 Q_1$:
\be
{1\over\Lambda^3} d^{cT} i\sigma_2 Q \chi_1^3 H_d
,\label{1.11}
\ee
[the discrete symmetry that should be
considered in this case is $Z_8\times Q_8$ (\ie\ $N=4$) (see table 1)],  
and the mass matrix eq.~(\ref{1.10a}) becomes
\be
M_d\simeq\pmatrix{
0 & \epsilon_1^3 & 0 \cr
-\epsilon_1^3 & \epsilon^2 & \epsilon\epsilon_\chi \cr
0 & \epsilon\epsilon_\chi & 1         \cr}
h_d\langle H_d\rangle
,\label{1.12}
\ee
where we understand that the coefficients are of order 1.
The texture structure of eq.~(\ref{1.12}) is similar to those considered 
in the literature \cite{texture}. 
We will assume that the CKM mixing matrix
and the quark masses can be recovered also in this case, although a
complete analysis including also the up-quark masses would be needed.
Even without performing the detailed analysis, we see that
the correct orders of magnitude can be obtained
if $\epsilon_1\sim\epsilon\simeq \lambda$,
which is the case if the $\uone_1$ and $\sutwo$ symmetries are broken at
the same scale.
Again, we find from eq.~(\ref{1.12}) that the left-handed
and the right-handed rotations are expected to be of the same order
of magnitude.

\vskip.5cm

As we have discussed, the quark mass matrix can be understood as arising
from horizontal symmetries in several ways. From a phenomenological point
of view, there is no way to distinguish amongst the different models that
lead to a satisfactory description of the quark masses.
As we will see, this arbitrariness is reduced
since the absence of large FCNC, which can be induced by 
the scalar partners of the quarks, favors a discrete
horizontal symmetry with the quark superfields 
in the $\mathbold{2+1}$ representation.

\section{The squark mass matrices}

We will  consider that the SSB terms are
the most general ones \cite{grisaru} allowed
 by the symmetry of the theory and  that they are generated at a high scale
$\approx M_P$.
If the first and second family superfields, $Q$, $D^c$ and $U^c$,
transform as a doublet under a horizontal symmetry $G_H$, the corresponding
squarks are degenerate.
 Nevertheless, the horizontal symmetry
 has to be broken to generate the masses of  the quarks  $u,c$ and $d,s$.
The $G_H$-breaking can affect the SSB parameters
and spoil the degeneracy of the squarks of the 
first two families \cite{savas}.
 In the ignorance of the underlying theory  above the scale $\Lambda$,
one can proceed as in the fermion sector and write
all  possible
non-renormalizable operators allowed by the symmetry
of the theory
that can contribute (when $G_H$ is broken) to the SSB terms.
This is explained in detail in the appendix.
These operators  determine the
  effective theory below $\Lambda$.
Of course, if we knew the underlying theory above $\Lambda$,
we could know which operators are in fact generated and which are not.
For example, if in the theory above $\Lambda$ the
 quark superfields
do not mix with other superfields,
then  no operator contributing to the squark soft masses
 (and involving
 superfields that break $G_H$)
 can be generated \cite{savas}.
In these theories
the super-GIM cancellation is guaranteed and
only charged currents can generate flavor-violating processes.
The non-renormalizable operators
contributing to the
fermion masses  could be  assumed to arise from Higgs  mixings.

Here we will adopt a more conservative approach.
Based on symmetry grounds, we will parametrize the scalar masses
in a model-independent way.
We will  assume the
 `onion'  flavor symmetry breaking
 suggested in the previous section for the fermions. This breaking pattern
allows us to write
the down-scalar masses
using an expansion in
 $\lambda$ (we neglect the SU(2)$_L\times$U(1)$_Y$
gauge contributions arising from the D-terms):
\be
{\bf M^2_{\tilde d}}=\pmatrix{
{\bf m^2_{Q}}+M_d^\dagger M_d          & {\bf m^{2\dagger}_{LR}}\cr
{\bf m^2_{LR}}& {\bf m^2_{D}}+M_d M_d^\dagger }\, ,
\label{squarkmass}
\ee
where\footnote{After  the first breaking step of eq.~(\ref{1.71}),
 contributions  [of $\calo(\lambda^2)$] are induced in
all entries of the
squark masses except for  those  of the
first family squarks since the latter
  transform non-trivially under the U(1)$_1$. Actually an
    $\calo(\lambda^2)$-contribution to the
11 entry
of ${\bf m^2_{Q,D}}$ is allowed by the U(1)$_1$, but
it can be absorbed in $m^2_{Q,D}$.
The rest of the
entries can only be generated after the  second breaking step
in  eq.~(\ref{1.71}).
If $G_H$ is gauged, there are extra contributions  coming from D-terms
that  we will comment on in the next section.}
\bea
{\bf m^2_{Q,D}}&\simeq& m^2_{Q,D}\pmatrix{
1         & \lambda^3        & \lambda^3 \cr
\lambda^3              & (1+\lambda^2)       & \lambda^2  \cr
\lambda^3 & \lambda^2        & m^2_{Q_3,D_3}/ m^2_{Q,D}\cr}\, ,
\label{squarkmassll}\\
{\bf m^2_{LR}}&\simeq& A_0m_b\pmatrix{
\lambda^4          & \lambda^3        & \lambda^3 \cr
\lambda^3              & \lambda^2       & \lambda^2  \cr
\lambda^3 & \lambda^2        & A_b/A_0\cr}+\mu M_d\tan\beta\, ,
\label{squarkmasslr}
\eea
where  $\mu$
is the Higgs mass in the superpotential.
We are only showing the order of
magnitude of  the  entries of the squark masses
based on symmetry considerations.
In a given  model, however,
the off-diagonal entries could be smaller
than those
in  eqs.~(\ref{squarkmassll}) and (\ref{squarkmasslr}).
For example, in our model I (see appendix),
 the squark mass matrix is of the form (\ref{squarkmass}) but
 in our model II the 12 and 13 entries of ${\bf m^2_{Q,D}}$ are of
order  $\lambda^5$ [see eq.~(\ref{a.5})].
As we will see later, however,
only the  diagonal entries are relevant  for phenomenological purposes.
We will  neglect the scale evolution of the different entries in
eq.~(\ref{squarkmass}),
 since
this effect is smaller than the uncertainty in
 the parameters of the model.
The soft mass matrix of the $U^c$-squark, ${\bf m^2_{U}}$, is  arbitrary
if the CKM matrix is assumed to come from the down sector.
Nevertheless,  the off-diagonal entries of ${\bf m^2_{U}}$
 cannot be larger than those
in eq.~(\ref{squarkmassll}).

Eq.~(\ref{squarkmassll}) can be diagonalized by a matrix $\widetilde V_{Q,D}$
of order
\be
\widetilde V_{Q,D}\simeq\pmatrix{
1         & \lambda        & \lambda^3 \cr
\lambda              & 1       & \lambda^2  \cr
\lambda^3 & \lambda^2        & 1}\, ,
\label{rotation}
\ee
that is of the order of $V_{\rm CKM}\sim U^d_{L,R}$.
We obtain the $Q$-squark soft masses:
\be
{\bf m^2_{Q}}\simeq
{\rm diag}(m^2_{Q},m^2_{Q}(1+\lambda^2),m^2_{Q_3})\, ,
\label{softmasses}
\ee
and similarly for $D^c$.
The horizontal symmetry only guarantees a degeneracy between
the first and second families up to $\calo(\lambda^2)\sim 4\times 10^{-2}$.
 As we will see in the next section, this is not
 enough
to suppress the supersymmetric
contribution to the $\varepsilon_K$ parameter if the CP-violating
phases are of $\calo(1)$.
One  possibility
 to further suppress this contribution will be to assume that
\be
 m^2_Q\gg m^2_{Q_3}\sim m^2_H\sim m^2_Z\, ,
\label{splitting}
\ee
and equivalently for the $U^c$ and $D^c$.
Notice that such a  splitting is not possible in a
scenario with universal soft masses but it is
possible here since we are considering
  general SSB terms.

Eq.~(\ref{splitting}) arises several questions.
First, one can wonder
how to generate such a splitting.
Second, it is not clear that eq.~(\ref{splitting})
is stable under radiative corrections.
It is known that supersymmetry  stabilizes the
scale  hierarchies
if the supersymmetry-breaking scale $\msusy$ is close to
the weak scale $m_Z$. If
 the supersymmetry-breaking masses of the first
 families are much  larger than  $m_Z$,
 one could be worried about whether such a splitting destabilizes
the weak scale.

In principle,
eq.~(\ref{splitting}) could be  a consequence of
the supersymmetry-breaking pattern at $M_P$. A large hierarchy in SSB
parameters is possible in string theories \cite{string}.
A second possibility is that eq.~(\ref{splitting})
is generated after integrating out some heavy states.
It is shown in refs.~\cite{murayama,pomarol,rattazzi} that, after integrating
out the heavy states, new contributions to the soft masses
of the light sparticles can
be induced.  Some of these contributions
 (see fig.~4d of ref.~\cite{pomarol})
 depend on  ratios of VEVs and masses of the
heavy states, and  could  be larger than $\msusy^2$.
It is crucial, in order to preserve the splitting (\ref{splitting})
under radiative corrections,
 to generate it
 at a scale below  $\Lambda$.
At this scale the $1^{st}$ and $2^{nd}$ families are
almost decoupled from the $3^{rd}$ family and the Higgs,
and the small soft masses of the latter
can be maintained naturally.
Nevertheless, the splitting
(\ref{splitting}) cannot be  very large since
the first and second families couple to the third one and to the Higgs
 through gauge interactions.
At the one-loop level,
there are D-term contributions
 to the renormalization group equations (RGEs)
 of the
Higgs $H_{u,d}$ and third family scalars given by
\be
\frac{dm^2_{i}}{dt}=\frac{3Y_{i}\alpha_1}{10\pi}
\left(m^2_Q-2m^2_U+m^2_D-m^2_L+m^2_E\right)\, ,
\ee
where
$Y_i$ is the hypercharge of the particle $i$,
$Y_i=(1,-1,1/3,2/3,-4/3,-1,2)$ for $i=(H_u,H_d,Q_3,D^c_3,U^c_3,L_3,E^c_3)$
where $L$ and $E^c$ denote the leptons
and  $t=\ln{\cal Q}$.
However, this contribution cancels if  the soft masses
respect an SU(5) symmetry,
 \ie\ $m^2_Q=m^2_U=m^2_E$ and $m^2_D=m^2_L$.
At the two-loop level,
the dominant contributions to the RGEs are given by \cite{jack}
\be
\frac{dm^2_i}{dt}=\beta^{(3)}_{m^2_i}+\beta^{(2)}_{m^2_i}+
\beta^{(1)}_{m^2_i}\, ,
\ee
where
\be
\beta^{(3)}_{m^2_i}=
\frac{2\alpha^2_3}{3\pi^2}
\left(2m^2_Q+m^2_U+m^2_D\right)\, ,
\label{color}
\ee
 for $i$ = color triplet,
\be
\beta^{(2)}_{m^2_i}=
\frac{3\alpha^2_2}{8\pi^2}
\left(3m^2_Q+m^2_L\right)\, ,
\ee
 for $i$ = weak doublet, and
\bea
\beta^{(1)}_{m^2_i}&=&
\frac{3Y^2_i\alpha^2_1}{200\pi^2}
\left(m^2_Q+3m^2_L+8m^2_U+2m^2_D+6m^2_E\right)\nonumber\\
  &+&\frac{3Y_i\alpha_1}{20\pi^2}\Biggl[\frac{8\alpha_3}
{3}\left(m^2_Q-2m^2_U+m^2_D\right)
+\frac{3\alpha_2}{2}\left(m^2_Q-m^2_L\right)\nonumber\\
&+&\frac{\alpha_1}{30}\left(m^2_Q-32m^2_U+4m^2_D
-9m^2_L+36m^2_E\right)\Biggr]\, ,
\eea
for $Y_i\not=0$.
Eq.~(\ref{color}) can induce a  large contribution
to the third-family-squark mass when
the  evolution from $M_G$ to $m_Q$ is taken into account:
\be
\Delta m^2_{Q_3,D_3,U_3}\simeq
\frac{4}{9\pi}\left[\alpha_3(M_G)-\alpha_3(m_Q)\right]
\left(2m^2_Q+m^2_U+m^2_D\right)\, .
\label{colorcon}
\ee
For equal $1^{st}$ and $2^{nd}$ family
soft masses, eq.~(\ref{colorcon})
 only allows for a splitting $m^2_Q/m^2_{Q_3}\simeq 30$.
 If only  $m^2_D$ dominates  (\ref{colorcon}),
the splitting can be $m^2_D/m^2_{Q_3}\simeq 120$.
Of course, there are other contributions to the RGEs of the third-family
squarks coming from the top Yukawa coupling and gluino
that will change  the estimate eq.~(\ref{colorcon})
---recent analysis can be  found also in ref.~\cite{giudice}.
Also the splitting (\ref{splitting}) could
be generated at a scale below $M_G$.
As a conservative value,  we will allow for a factor 10
 splitting
 between the soft masses of the
$1^{st}-2^{nd}$ families
and the $3^{rd}$ one.

\section{Phenomenological implications}

\subsection{FCNC processes and EDMs}

We will work in the basis in which the squark mass
matrices ${\bf m^2_{Q}}$,  ${\bf m^2_{D}}$ and  ${\bf m^2_{U}}$ are diagonal.
We can go to this basis through a
superfield rotation $\widetilde V_{Q,D,U}$ that will
redefine
 the quark mass matrices as
\bea
M_d&\rightarrow&\widetilde V^T_{D}M_d\widetilde V_{Q}=
\widetilde V^T_{D}U^{d}_RM^{diag}_dU^{d\dagger}_L\widetilde V_{Q}\, ,
\nonumber\\
M_u&\rightarrow&\widetilde V^T_{U}M_u\widetilde V_{Q}=
\widetilde V^T_{U}U^{u}_RM^{diag}_uU^{u\dagger}_L\widetilde V_{Q}\, ,
\eea
and similarly for the left--right mixing mass matrices ${\bf m^2_{LR}}$.
Hence all the flavor mixings will arise  from the four matrices
\bea
V^Q&\equiv& U^{d\dagger}_L\widetilde V_{Q}\, ,\ \ \ \ \ \ \
\overline{V}^Q\equiv U^{u\dagger}_L\widetilde V_{Q}\, ,\nonumber\\
V^D&\equiv& U^{d\dagger}_R\widetilde V^{*}_{D}\, ,\ \ \ \ \ \ \
V^U\equiv U^{u\dagger}_R\widetilde V^*_{U}\, ,
\label{matrices}
\eea
and ${\bf m^2_{LR}}$ that is not diagonal in the above basis.
The  $\lambda$-dependence of the matrices (\ref{matrices})
is easily derived from
  (\ref{rotation}) and the assumptions  $U^d_{L,R}\sim V_{\rm CKM}$
and $U^u_{R}\sim V_{\rm CKM}(\lambda\rightarrow \lambda_u)$:
\be
V^{Q,D}\sim\overline{V}^Q\simeq\pmatrix{
1         & \lambda        & \lambda^3 \cr
\lambda              & 1       & \lambda^2  \cr
\lambda^3 & \lambda^2        & 1}\, ,
\label{flavormatrix}
\ee
and similarly  for $V^U$ with $\lambda\rightarrow\lambda_u$.
Had we taken the off-diagonal entries
of ${\bf m^2_{Q,D}}$ smaller than those in (\ref{squarkmassll}),
as in our model II, the matrices $V^{Q,D}$ 
would have been of the same order of (\ref{flavormatrix}),
since they
would still have been originated from   $U^d_{L,R}$.

Let us start considering the contributions to FCNC processes
and EDMs induced by the
$1^{st}$ and $2^{nd}$ family squarks.
The most stringent constraints on the model come from
gluino-mediated contributions to CP-violating
 observables.
The contribution to the CP-violating
$\varepsilon_K$ parameter of the $K$--$\overline{K}$ system
is dominated by
diagrams involving $Q$ and $D^c$ squarks in the same loop \cite{FCNC,hagelin}.
If $m^2_{Q}\simeq m^2_{D}\gg m^2_{\tilde g}$, where  $m_{\tilde g}$
is  the gluino  mass, we have
\cite{hagelin}:
\be
\varepsilon_K\simeq \frac{\alpha^2_3}{54\sqrt{2}}
\frac{f^2_Km_K}{m^2_Q\Delta m_K}\left[1+\frac{2}{3}
\left(\frac{m_K}{m_s+m_d}\right)^2\right]
{\rm Im}\left\{\frac{V^Q_{11}{\delta m^2_{ Q}}V^{Q*}_{21}}{m^2_{ Q}}
\frac{V^D_{11}{\delta m^2_{ D}}V^{D*}_{21}}{m^2_{ D}}\right\}\, ,
\label{epsilon}
\ee
where  $\delta m^2_{Q,D}$ is the mass-squared  difference between the
 first and second
families of squarks. Note that we have CP violation even if
 only two families
are considered \cite{nirb}.
Using the experimental value $\varepsilon_K\simeq 2.3\times 10^{-3}$,
we get the constraint
\be
\left(\frac{1\ {\rm TeV}}{m_Q}\right)^2\left|
\frac{V^{Q}_{11}{\delta m^2_{ Q}}V^{Q}_{21}}{m^2_{ Q}}
\frac{V^{D}_{11}{\delta m^2_{ D}}V^{D}_{21}}{m^2_{ D}}\right|
\sin \varphi<  1.2\times 10^{-6}\, ,
\label{bound}
\ee
where
$\varphi={\rm Arg}(V^{Q}_{11}V^{Q*}_{21}
V^{D}_{11}V^{D*}_{21})$. From
eqs.~(\ref{softmasses}) and (\ref{flavormatrix}),  we obtain
\be
\left|\frac{V^{Q,D}_{11}{\delta m^2_{ Q,D}}V^{Q,D}_{21}}{m^2_{ Q,D}}\right|
\simeq \lambda^3\sim 8\times  10^{-3}
\ \ \Rightarrow\ \ m_{Q,D}>5\ {\rm TeV}
\sqrt{\frac{\sin \varphi}{1/2}}\, .
\ee
We  then see that
an approximate horizontal
symmetry under which the first two  families
transform as a doublet cannot guarantee a small
 contribution to $\varepsilon_K$.
One  is forced to have large
squark masses {\it or } small CP-violating phases.
If the horizontal symmetry is gauged,
extra gauge contributions to the squark masses arise when the symmetry
 is broken:
\be
{\bf \Delta m^2_{Q,D}}\simeq T_Am^2_A\,  ,
\label{dterm}
\ee
where $T_A$ are the generators of $G_H$
and $m^2_A$ are SSB parameters of the order
 of the soft masses of the scalars that break  $T_A$.
Such contributions induce a $\delta m^2_{Q,D}\sim m^2_A$
that for $m^2_A\sim m^2_{Q,D}$ spoil the super-GIM cancellation.
Hence a gauged $G_H$ will only be allowed if
 $m^2_{Q,D}$ are much larger than the other soft masses.

Contributions to the neutron  EDM can also be important, in particular
due to  one-loop gluino diagrams.
The induced down-EDM for $m^2_{Q}\sim m^2_{D}\gg m^2_{\tilde g}$
 is given by
\cite{buchmuller}
\be
d_d=\frac{e\alpha_3}{9\pi m^4_Q}
{\rm Im}\big\{ m^*_{\tilde g}(V^{D}{\bf m^2_{LR}}V^{Q\dagger})_{11}\big\}\, .
\label{edmq}
\ee
If the phase in   $ m_{\tilde g}$ or in the
entry  $(V^{D}{\bf m^2_{LR}}V^{Q\dagger})_{11}$ is  of
order 1, eq.~(\ref{edmq}) gives a too large
 contribution to
 the neutron EDM [$d_n\simeq (4d_d-d_u)/3$ and  $|d_n|_{exp}<10^{-25}e\, $cm],
unless
 the squark masses are above the
TeV-scale.
Notice that the phase in
 $(V^{D}{\bf m^2_{LR}}V^{Q\dagger})_{11}$  is related to
those in  $V^{Q,D}$ and then has
 the same origin that the
 CKM  phase,  which is known to be  of $\calo(1)$.

The above constraints suggest the following possible scenarios\footnote
{Scenarios similar to (a) and (b) were also suggested in ref.~\cite{dine}.}:

\begin{quote}
 \hspace{-.6cm}(a)
The squark masses of the
first and second families are larger than
the other soft masses\footnote{
Actually,  eq.~(\ref{bound}) and the down-EDM
eq.~(\ref{edmq}) can be suppressed by only increasing
 $m^2_Q$ or $m^2_D$.
This suggests an alternative
   scenario where only $m^2_D$
is large (in an SU(5) GUT also $m^2_L$ would be large).},
 $m^2_i\sim 10^2\msusy^2$.
As we already
 mentioned, this does not lead to a   naturalness problem, since
these two families are almost decoupled from the Higgs.
This scenario allows for  phases in the SSB terms of $\calo(1)$ and
for gauged horizontal symmetries.

 \hspace{-.6cm}(b)
All the CP-violating phases
are small, $\varphi\sim 10^{-2}$
 (we now denote by $\varphi$ a generic CP-violating phase).
This is technically natural, since the phases are protected by an extra
symmetry, \ie\   the CP-symmetry.
This possibility can
 arise if
 CP is broken spontaneously.
In this case, small complex parameters could be generated
in the same way that one generates the small Yukawa couplings (see section 2).

 \hspace{-.6cm}(c)
The phases
in $(V^{Q}_{11}V^{Q*}_{21}V^{D}_{11}V^{D*}_{21})$
and  $m^*_{\tilde g}(V^{D}{\bf m^2_{LR}}V^{Q\dagger})_{11}$
can be rotated away
(\ie\ they are not physical).
As we will see,
the first condition  is satisfied in models with
special texture zeros in the fermion
mass matrices such as our model II.
The second condition is fulfilled if
supersymmetry is
assumed to be broken by a hidden sector
such that the SSB terms are real.
\end{quote}

In  scenario (a),
contributions to FCNC processes and EDMs (of the first two families) can
arise from one-loop diagrams involving
the  third-family squarks \cite{hkr,bh}.
Even that  they are
 suppressed because of the small
 mixing between  the third and the lightest families,
  they can be close
 to the experimental values.
Stop-loops can generate a large
 up-EDM:
\be
d_u=\frac{2e\alpha_3}{9\pi m_{\tilde g}}|V^U_{13}\overline{V}^Q_{13}|
\sin\varphi\sin 2\theta_t
\left[x_1I(x_1)-x_2I(x_2)\right]\, ,
\label{edmup}
\ee
where
$x_i=m^2_{\tilde g}/m^2_{\tilde t_i}$ and
$I(x)=[(1+x)/2+{x}\ln x/({1-x})]/{(1-x)^2}$;
the angle
 $\sin\theta_t$ defines the  left--right stop mixing,
 \ie\ $\tilde t_1=-\sin\theta_t\tilde t_R+\cos\theta_t\tilde t_L$.
Taking
$V^U_{13}\sim \lambda^3_u\sim 2\times 10^{-4}$,
and $(\sin\varphi\sin 2\theta_t)\sim 0.25$,
the contribution (\ref{edmup})
to the neutron EDM is close to the experimental value for
$m_{\tilde t_2}\gg  m_{\tilde g}\sim 100\ {\rm GeV}\gg m_{\tilde t_1}$.

Larger effects arise from the
 bottom sector. Constraints on the sbottom masses can be obtained
from
$\varepsilon_K$. Taking
 $m^2_{Q_3}\simeq m^2_{D_3}$, we obtain
\cite{hagelin}:
\be
\left(\frac{1\ {\rm TeV}}{m_{Q_3}}\right)^2
\big|V^Q_{13}V^Q_{23}V^D_{13}V^D_{23}\big|\sin\varphi< 10^{-7}\times
\left\{\begin{array}{ll}
4&\ {\rm for}\  m^2_{Q_3}\gg m^2_{\tilde g}\\
1.6&\ {\rm for}\  m^2_{Q_3}\simeq m^2_{\tilde g}\end{array}\right.\, ,
\label{boundb}
\ee
which implies (since $V^{Q,D}_{13}V^{Q,D}_{23}\sim\lambda^5$)
\be
 m_{Q_3,D_3}>
\sqrt{\frac{\sin\varphi}{1/2}}\times
\left\{\begin{array}{ll}
350\ {\rm GeV}&\ {\rm for}\  m^2_{Q_3}\gg m^2_{\tilde g}\\
550\ {\rm GeV}&\ {\rm for}\  m^2_{Q_3}\simeq m^2_{\tilde g}\end{array}\right.
\label{boundnew}
\, .
\ee
The contribution to  the down-EDM is given by
(for $m_{Q_3}^2\simeq m_{D_3}^2$)
\be
d_d=\frac{e\alpha_3m_b}{9\pi m^4_{Q_3}}
\big|V^D_{13}V^Q_{13}\big|\sin\varphi\, m_{\tilde g}
({A_b+\mu\tan\beta})\times
\left\{\begin{array}{ll}
1&\ {\rm for}\  m^2_{Q_3}\gg m^2_{\tilde g}\\
1/6&\ {\rm for}\  m^2_{Q_3}\simeq m^2_{\tilde g}\end{array}\right.
\, ,
\label{edmdown}
\ee
which can put   stronger constraints if  $\tan\beta$ is large.
Notice that
the  constraints eqs.~(\ref{boundb}) and  (\ref{edmdown})
can be relaxed if either $m_{Q_3}$ or  $m_{D_3}$ are
 increased. Since $D^c_3$ is decoupled from  $H_u$
at tree level, $m_{D_3}$
 could be  larger than $m_{H_u}\sim m_Z$ without an extreme fine-tuning.
Constraints on only $m_{Q_3}$
 can also be  obtained from $\varepsilon_K$ (or $\Delta m_K$)
but they are about
an order of magnitude  smaller than eq.~(\ref{boundnew}) \cite{hagelin}.

Contributions to $\Delta m_{B_d}$ and $b\rightarrow s\gamma$
can also be important.
They
are, however, always
 suppressed by  a mixing factor
of order $\lambda^3\sim 8\times 10^{-3}$
and  $\lambda^2\sim 4\times 10^{-2}$ respectively.
Several diagrams contribute to
these processes
(involving  chargino--stop, gluino--sbottom, Higgs--top and W--top loops).
They  give similar contributions with an arbitrary
 relative sign \cite{bertolini},
rendering it difficult to get predictions for these processes.

In scenario (b), the $3^{rd}$ family contributions  to $\varepsilon_K$
and EDMs are very small since they are suppressed by the small
CP-violating phases.
 In fact, also the
usual SM contribution to $\varepsilon_K$ arising from a box diagram
involving a $W$ is small;
 $\varepsilon_K$
is dominated by diagrams involving  first- and second-family squarks
 and gluinos [eq.~(\ref{epsilon})],
and we have to rely on these contributions to explain
the experimental value of $\varepsilon_K$.
Even if $\varphi$ is small,
constraints on the squark masses could arise from  $\Delta m_K$,
but they are around an order of magnitude smaller than those
from $\varepsilon_K$ \cite{hagelin}.

Let us finally consider scenario (c). The phases
of the $ij$ entries  $(i,j=1,2)$ of  $V^{Q}$ and $V^D$
 can be rotated away
 in models where (in the basis
where ${\bf m^2_{Q,D}}$ are diagonal)
 the 11 entry of $M_d$ is zero.
This is because in a two-family model one has the freedom to redefine
the phases in the down quarks such that  only
 the 11 entry  of $M_d$ is complex.
An example where $(M_d)_{11}=0$ is our model II [see eq.~(\ref{1.12})].
The condition
${\rm Im}\big\{ m^*_{\tilde g}(V^{D}{\bf m^2_{LR}}V^{Q\dagger})_{11}\big\}=0$
requires, first,
that
supersymmetry is
 broken by a hidden sector.
In this case (see appendix),
 ${\bf m^2_{LR}}\simeq (A_0+\mu\tan\beta)M_d$, which  implies
${\rm Im}\big\{ m^*_{\tilde g}(V^{D}{\bf m^2_{LR}}
V^{Q\dagger})_{11}\big\}\simeq
m_d\, {\rm Im}\{m^*_{\tilde g}(A_0+\mu\tan\beta)\}$.
If
 the SSB parameters are real (as suggested in certain theories \cite{choi}),
 the  contributions to the EDMs vanish.
Of course, this does not mean that all the physical CP-violating phases
are zero. When the third family is considered,
there are phases that cannot be rotated away such as the CKM phase.

\subsection{$\mu\rightarrow e\gamma$ and proton decay in GUTs}

In GUTs such as SU(5) or SO(10), the lepton Yukawa matrix
 is  related to the quark Yukawa matrix.
If our model is embedded in such GUTs,
 the lepton  and sleptons mass matrices
will be of the form  of eqs.~(\ref{1.3}) and (\ref{squarkmass}), respectively,
but with   different Clebsch factors \cite{adhrs}  acting on
the different entries.
Such Clebsch factors depend on the GUT and they are
usually a number between  $0.1$--$10$.
The lepton number is  violated
and flavor-violating  processes
such as $\mu\rightarrow e\gamma$ can be induced.
In scenario (a), only the contributions from
the stau are relevant \cite{bh}. From the experimental value,
${\rm BR}(\mu\rightarrow e\gamma)|_{exp}<4.9\times 10^{-11}$,
one has, taking $m^2_{L_3}\simeq m^2_{E_3}$,
\be
\left(\frac{100\ {\rm GeV}}{m_{L_3}}\right)^2
\left[\frac{m_{\tilde B}m_\tau(A_\tau+\mu\tan\beta)}
{m^2_{L_3}m_\mu}\right]V^L_{13}V^E_{23}
<10^{-3}\times
\left\{\begin{array}{ll}
1/2&\ {\rm for}\  m^2_{L_3}\gg m^2_{\tilde B}\\
3&\ {\rm for}\  m^2_{L_3}\simeq m^2_{\tilde B}\end{array}\right.\, ,
\label{constraint}
\ee
where $V^{L,E}$ define
 the rotations that diagonalize the lepton mass matrix $M_l$
(in the basis where ${\bf m^2_{L,E}}$ are diagonal)
and $m_{\tilde B}$ is the mass of the bino that is assumed to be
a mass-eigenstate.
Taking  $V^{L,E}\sim V^Q$,
eq.~(\ref{constraint}) can be
 satisfied for $m_{L_3}\sim m_{\tilde B}\sim(A_\tau+\mu\tan\beta)\sim
150$ GeV.
The above constraint comes
from the one-loop diagram
involving a  left--right stau mixing
and again can be loosened if either  $m_{L_3}$ or  $m_{E_3}$
is large.
Bounds on only $m_{L_3}$ or  $m_{E_3}$
are about an order of magnitude weaker.
In scenarios (b) and (c) there are also contributions to
 $\mu\rightarrow e\gamma$ from diagrams involving the first- and
second-family squarks. Particularly dangerous contributions are
those involving a left--right mixing. They  lead to the constraint
(for $m^2_{L}\simeq m^2_{E}$)
\be
\left(\frac{100\ {\rm GeV}}{m_{L}}\right)^2
\left[\frac{m_{\tilde B}\left(V^E{\bf m^2_{LR}}V^{L\dagger}\right)_{12}}
{m^2_{L}m_\mu}\right]
<10^{-3}\times
\left\{\begin{array}{ll}
1/2&\ {\rm for}\  m^2_{L}\gg m^2_{\tilde B}\\
3&\ {\rm for}\  m^2_{L}\simeq m^2_{\tilde B}\end{array}\right.\, ,
\label{constraintb}
\ee
where   ${\bf m^2_{LR}}$ now denotes the left--right mixing
in the selectron and smuon sector.
For general SSB terms (see appendix),  ${\bf m^2_{LR}}$ is not
proportional to $M_l$ and then $V^E{\bf m^2_{LR}}V^{L\dagger}$
is non-diagonal. The 12 entry  of the latter
is expected to be $\sim A_0\lambda^3m_\tau$,
 leading to the bound
\be
m_{L}>700\ {\rm GeV}\ \ {\rm for}\
 m_{L}\simeq A_0\simeq m_{\tilde B}\, .
\ee
If supersymmetry is assumed to be broken by a hidden sector,
one has for a two-family model (see appendix) that
 $V^E{\bf m^2_{LR}}V^{L\dagger}\simeq M^{diag}_l(A_0+\mu\tan\beta)$
and the 12 entry is zero
up to $\calo(\lambda^5)$. In this case, the constraint (\ref{constraintb})
can be easily accommodated.

It should be kept in mind, however, that
the above bounds present a large uncertainty due to the
 unknown Clebsch factors that relate $V^{L,E}$ with $V^{Q,D}$.

Another feature in
 SUSY GUTs is the
proton decay that arises from  dimension-five operators induced
by colored Higgsinos.
The proton decay  rate depends not only on the coupling
of the colored Higgsinos to the quarks and leptons (which
depends on Clebsch factors) but on the squark and slepton
masses as well \cite{weinberg}.
For universal soft masses,
the
dominant proton decay modes
are \cite{protondecay}
 $p\rightarrow K^+\nu_{e,\mu}$
 and $p\rightarrow \pi^+\nu_{e,\mu}$; they
 arise from
 loop diagrams involving the  scalars of first two families
and the $\widetilde W$.
In scenario (a) such  diagrams are suppressed
since  $m^2_i\gg m^2_{\widetilde W}$  for $i=1^{st}-2^{nd}$ family
scalars.
Nevertheless, diagrams involving only the third-family sparticles
can also contribute to proton decay.
The dominant mode is now\footnote{
Contributions to $p\rightarrow K^+\nu_{e}(\nu_\mu)$
  are extra suppressed by the  small mixing of
the first (second)  with the third family.}
 $p\rightarrow K^+\nu_\tau$.
It is important to notice that, unlike the
case with universal soft masses,
in  scenario (a)
 either the gluino or the wino exchange contribute to this decay mode.

Gluino exchange diagrams can also contribute to  $p\rightarrow K^0\mu$
if there is a large splitting between the squark masses.
This decay could then be used to probe scenario (a).
Note, however, that  $p\rightarrow K^0\mu$
needs
an extra mixing factor between second and third family
that is not necessary in
  the decay
  $p\rightarrow K^+\nu_{\tau}$.

\subsection{CP-violating phases in the SSB terms}

In  scenario  (a),
 the SSB parameters (and the Yukawas) are allowed to have phases
of $\calo(1)$
giving rise to a  phenomenology richer than that in the MSSM
with real SSB parameters.
Charginos, neutralinos and stops can have complex mass matrices and
their interactions will  violate CP.
It is interesting to analyse
the  effects
 on the upper bound on the
lightest-Higgs mass.
If  phases in the SSB terms are allowed,
the parameters of the
Higgs potential can be complex. It is possible, however,
 to rotate the phases away from the tree-level potential.
CP-violating phases will be present in
 the stop left--right mixing
\be
m^2_{LR}\equiv
m_t\left(A_te^{i\varphi_t}+\mu e^{i\varphi_{\mu}}\cot\beta\right)\, ,
\label{lrmixing}
\ee
that, at the one-loop level, will affect the Higgs mass.
It can be shown\footnote{We thank H. Haber for his collaboration on
this point.} that the upper bound on
the mass of the lightest Higgs, written
as a function of $|m^2_{LR}|$, is not
modified by the phases in eq.~(\ref{lrmixing}). Only
 the relative phase between $\mu$ and $A_t$
can affect  $|m^2_{LR}|$, and consequently
modify the Higgs mass bound.

There are also important  effects of the CP-violating phases
 on the dark-matter density of the  lightest supersymmetric particle (LSP).
 It has been shown in ref.~\cite{olive} that when phases are included
in the sparticle mass matrices, the limits on the LSP mass can be
relaxed by a factor $2$--$3$.

In general, CP-violating phases change the relation between
masses and
interactions of the sparticles and then modify the
 experimental bounds from direct detection.

In scenario (b) all the CP-violating phases are
small, $\varphi\sim\calo(10^{-2})$, and the above effects are not present.
 The best place to probe this scenario
 will be in future $B$-factories.
Since the  CP-asymmetries in  $B$ decays are
 proportional to the CP-violating phases in the
CKM matrix, one expects  much smaller asymmetries in these models than
in  models with $\varphi\sim \calo(1)$.

\section{Conclusions}

We have discussed how the observed quark masses and CKM mixings
 point to a
definite flavor structure for the down-quark mass matrix in the gauge
eigenstates, that can be described in terms of an approximate horizontal
symmetry $G_H$ for the SM. Assuming that $G_H$ is realized at some
superheavy scale and commutes with some GUT group such as SO(10), 
and that it leaves invariant the light Higgs field, we have
shown that it should be a subgroup (either continuous or discrete)
of $\utwo\times Z_2$, characterized by the symmetry-breaking pattern
$\utwo\to\uone_1\to 1$. This class of non-Abelian groups has a
two-dimensional representation, so that the quark families can belong to the
$\mathbold{2}+\mathbold{1}$ representation; this choice is favored
in the supersymmetric version of the theory since it allows for a super-GIM
mechanism. We have provided explicitly two models based on $\utwo\times Z_2$
(or on one of its discrete subgroups). Model I predicts
the correct orders of
magnitude for the entries of the mass matrix, leading to the
most general structure consistent with the observations; in model II,
on the contrary,
texture zeros are generated so that a more predictive form of the mass matrix
is generated.

In the scalar sector,
the horizontal symmetry forces the  squarks of the first two families
 to be degenerate. Nevertheless,
the breaking of $G_H$ spoils such  degeneracy
unless the quark superfields do not mix with other heavy superfields
\cite{savas}.
The breaking of the degeneracy in the
 down sector is expected to be
\be
\frac{\delta m^2_{Q}}{m^2_{Q}}\simeq\frac{ m_s}{m_b}\, ,
\ee
which is sufficient to suppress the squark contributions to FCNC
processes except for those to $\varepsilon_K$.
We have proposed three possibilities that lead to a
 $\varepsilon_K$ in agreement with the data:
(a)  the first- and second-family
 scalars are heavier ($\sim$ few TeV)
than the other scalars ($\sim$ 100 GeV); (b)  all CP-violating
phases are small, $\varphi\sim\calo(10^{-2})$. (Both possibilities
provide also a solution  to the  EDM problem \cite{buchmuller}
in the MSSM.)
A third possibility (c) is to have a down-quark
 mass matrix with  specific texture zeros such that the dominant
contribution to   $\varepsilon_K$ vanishes. We have shown an example
where this scenario is realized.

The phenomenology of these scenarios is very different from that
in the MSSM with universal soft masses.
In scenario (a), we have shown that
 the third-family squarks  can  give important
contributions to FCNC processes
and EDMs.  For sbottom masses around 300 GeV, these processes are
predicted to be around the experimental values.
Scenario (a) also leads to a very different phenomenology
in the context of GUTs. The  proton decay and  $\mu\rightarrow e\gamma$
are induced by the third-family scalars and then
they depend on the mixing angles of the first and second family with
the third one.
The proton decay mode  $p\rightarrow K^0\mu$ is enhanced
(compared to the case with universal soft masses) and
could test this scenario.

Scenario (a) allows for phases of order 1
 in the soft terms. These phases can have implications
in the Higgs mass, dark-matter density and $B$ decays.

The phenomenology of scenarios (b) and (c)
is more similar to that
in the MSSM with universal soft masses.
These scenarios could be distinguished, however,
  in the future $B$-factories where the CP-violating phases
in the $B$-decays
will be measured.

\vskip .2in
\noindent {\large{\bf Acknowledgements}}

\noindent
We thank S. Dimopoulos, B. de Carlos, H. Haber, 
G. Giudice and N. Rius for discussions.

\newpage


\noindent
{\Large{\bf Appendix. $G_H$-breaking effects in the SSB terms}}

\vskip.5 cm

\noindent
To analyse how
 the $G_H$-breaking affects the SSB terms,
it is convenient to work within the superfield formalism.
Softly broken supersymmetric theories can be formulated in the superfield
formalism by using a spurion external field, $\eta$ \cite{grisaru}.
 Supersymmetry
is broken by giving to this superfield a $\theta$-dependent value,
$\eta\equiv\msusy\theta^2$.
We will follow  ref.~\cite{pomarol} but
 extending that analysis to include non-renormalizable operators.
The SSB terms for the quark superfield ${\bf Q}=(Q,Q_3)^T$ can be written by
\bea
{\cal L}_{soft}&=&\int d^4\theta  {\bf Q}^{\dagger}
\biggl[\bar\eta\Gamma_Q(\phi)^\dagger+\eta\Gamma_Q(\phi)
-\bar\eta\eta \{Z_Q(\phi)-
\Gamma_Q(\phi)^\dagger\Gamma_Q(\phi)\}\biggr]{\bf Q}\nonumber\\
&-&\left(\int d^2\theta \eta W'({\bf Q},\phi)+h.c.\right)\, ,
\label{softsf}
\eea
and similarly for the other superfields;
   $\Gamma_Q(\phi)$ and $Z_Q(\phi)$ are matrix operators made
of the superfields $\phi$
that break $G_H$.
When the $\phi$ get VEVs, such matrices contribute
to the SSB parameters of ${\bf Q}$.
Explicitly, the contribution to the trilinears and
soft masses can be obtained  by writing
eq.~(\ref{softsf}) in  component fields [and after
replacing $\eta(\bar\eta$) by $\msusy\theta^2(\bar\theta^2$)]:
\begin{eqnarray}
A_{D_iQ_jH_d}&=&\msusy\left[Y_{ijH_d}'+Y_{ljH_d}\langle\Gamma_{D}\rangle_{li}
+Y_{ilH_d}\langle\Gamma_{Q}\rangle_{lj}
+Y_{ijH_d}\langle\Gamma_{H_d}\rangle\right]\, ,\nonumber\\
{\bf m^2_Q}&=&\msusy^2\langle Z_Q\rangle\, ,
\label{SSBcomp}
\end{eqnarray}
where
\be
 Y_{ijH_d}\equiv\frac{\partial W_{eff}}{\partial D^c_i\partial Q_j
\partial H_d}\, ,
\ee
 $W_{eff}$  being the low-energy effective superpotential;
 similarly for
 $Y'_{ijH_d}$, but replacing $W_{eff}$ by
 $W'_{eff}\equiv W'({\bf Q},\langle\phi\rangle)$.
The  $W'$
is a general holomorphic function of the
 superfields that is in principle
different from the superpotential $W$. However,
in  theories
where supersymmetry is broken by a hidden sector that does
not  couple to the observable sector in the superpotential,
one has that ($W$-proportionality condition \cite{pomarol})
\be
W'=aW\, ,
\label{proport}
\ee
where $a$ is a constant. Eq.~(\ref{proport}) is not modified either by
higher-order corrections (by the non-renormalization theorem) or
by integrating out heavy superfields \cite{pomarol}.

To analyse the  contributions to the soft masses,
 we write all   non-renormalizable operators
 ${\bf Q}^\dagger Z(\phi)\eta\bar\eta{\bf Q}$
allowed by $G_H$ and the MSSM group
in the effective theory below $\Lambda$.
We are assuming that $G_H$ is not gauged ---otherwise
there are additional contributions, eq.~(\ref{dterm}).
The $G_H$-invariant renormalizable operator is given by
(hereafter we will omit the $\eta$ superfield)
\be
{\bf Q}^\dagger\pmatrix{1& & \cr &1&
\cr & &b}{\bf Q}\, ,
\label{a.1}
\ee
which leads to  equal soft masses for the
  first two families.
Additional non-renormalizable operators involving
 $\Phi$ and $\chi$ (see table 1) are
\be
{\bf Q}^\dagger
\pmatrix{
 \Phi^*\Phi^T+i\sigma_2\Phi\Phi^\dagger i\sigma_2&
 \Phi^*\chi \cr
\chi^\dagger \Phi^T &
0 \cr}{\bf Q}\, ,
\label{a.2}
\ee
where for each term we understand a coefficient
 divided by the scale $\Lambda^2$,
and we omit
terms proportional to the operators of eq.~(\ref{a.1}).
The operators in eqs.~(\ref{a.1}) and
(\ref{a.2}) are common to the both models I
and II. The  specific operators to each model are given below.

{\bf Model I}: The addition of the field $\Phi'$ gives rise to
  operators similar to those in eq.~(\ref{a.2}) since
$\Phi$ and $\Phi'$ belong to the same representation of $G_H$.
When the fields $\Phi$,  $\chi$ and $\Phi'$ get VEVs
 [eqs.~(\ref{1.8}), (\ref{1.8c}) and (\ref{1.8b})],
 soft mass matrices of the form of
eq.~(\ref{squarkmassll}) are generated
if we identify $\epsilon\sim\lambda$.

{\bf Model II}: The field $\chi_1$ gives rise to the following invariants
\be
{\bf Q}^\dagger
\pmatrix{ i\sigma_2\Phi\Phi^T(\chi_1^{3})^\dagger&
i\sigma_2\Phi\chi(\chi_1^{3})^\dagger  \cr
-\chi_1^{3}\chi^\dagger\Phi^\dagger i\sigma_2&
0 \cr}{\bf Q}\, .
\label{a.4}
\ee
Then, from eqs.~(\ref{1.8}), (\ref{1.8c}) and
(\ref{1.13}),  and the identification $\epsilon\sim\lambda$,
we find
\be
{\bf m^2_Q}\simeq m_Q^2\pmatrix{
1         & \lambda^5   & \lambda^5 \cr
\lambda^5 & (1+\lambda^2) & \lambda^2 \cr
\lambda^5 & \lambda^2   & m^2_{Q_3}/m^2_{Q} \cr}
,
\label{a.5}
\ee
where we have absorbed in the parameter $m_Q^2$
a ${\cal O}(\lambda^2)$-contribution to the 11 entry.

The contributions to the trilinear SSB terms are more subtle.
If the $W$-proportionality condition (\ref{proport})  does not hold,
the effective terms induced in $W$ will also be induced in $W'$, but
with different coefficients, \ie\ $Y_{ijH_d}\sim Y'_{ijH_d}$.
Then the $\lambda$-dependence of
$({\bf m^2_{LR}})_{ij}=A_{D_iQ_jH_d}
\langle H_d\rangle+\mu (M_d)_{ij}\tan\beta$
 will be  the same as that of $(M_d)_{ij}=Y_{ijH_d}\langle H_d\rangle$.
Nevertheless, since
 they are not exactly equal, they  cannot be
diagonalized  and made  real by the same rotation,  and then
flavor-violating processes and EDMs can be induced.

If the condition (\ref{proport})  holds, one has from eq.~(\ref{SSBcomp})
\be
{\bf m^2_{LR}}=\msusy[aM_d+\langle\Gamma^T_D\rangle
 M_d+M_d\langle\Gamma_Q\rangle+M_d\langle\Gamma_{H_d}\rangle]
+\mu M_d\tan\beta\, ,
\ee
and the breaking of the proportionality ${\bf m^2_{LR}}\propto M_d$
 only arises from the operators
${\bf Q}^\dagger \eta\Gamma_Q {\bf Q}$ and
 ${\bf D}^{c\dagger} \eta\Gamma_D {\bf D}^c$ that
 are of the
same type as those in  eqs.~(\ref{a.1})--(\ref{a.4}).
Hence,  if only
 two families are considered, one has in  model I that
 $\langle\Gamma_{Q,D}\rangle=
\langle\Gamma^{diag}_{Q,D}\rangle\identity+\calo(\lambda^2)$ and
\be
{\bf m^2_{LR}}\simeq
 (A_0+\mu\tan\beta)M_d+\msusy\, m_{b}
\pmatrix{\lambda^6&\lambda^5\cr\lambda^5&\lambda^4\cr}\, ,
\ee
where $A_0=\msusy[a+\langle\Gamma_{H_d}\rangle
+\langle\Gamma_Q^{diag}\rangle+\langle\Gamma_D^{diag}\rangle]$.

\vskip .5in
\end{document}